\documentclass[prl,showpacs,twocolumn,superscriptaddress]{revtex4-1}

\usepackage{amsfonts,amsmath,amssymb,graphicx}
\usepackage[usenames,dvipsnames]{xcolor}
\usepackage{pdfpages}

\newcommand{\Exp}[1]{e^{#1}}

\newcommand{\ket}[1]{|#1\rangle}
\newcommand{\moy}[1]{\langle#1\rangle}

\newcommand{\out}{\mathrm{out}}
\newcommand{\SNR}{\mathrm{SNR}}

\newcommand{\sx}{\hat{\sigma}_x}
\newcommand{\sz}{\hat{\sigma}_z}
\newcommand{\sxj}{\hat{\sigma}_{xj}}
\newcommand{\szj}{\hat{\sigma}_{zj}}

\newcommand{\hAa}{\hat{a}^\dag\hat{a}}
\newcommand{\gz}{g_\mathrm{z}}
\newcommand{\gzj}{g_{\mathrm{z}j}}
\newcommand{\gx}{g_\mathrm{x}}
\newcommand{\gxj}{g_{\mathrm{x}j}}
\newcommand{\tgz}{\tilde{g}_\mathrm{z}}
\newcommand{\bgz}{\bar{g}_\mathrm{z}}
\newcommand{\chiz}{\chi_\mathrm{z}}
\newcommand{\wc}{\omega_\mathrm{r}}
\newcommand{\wa}{\omega_\mathrm{a}}
\newcommand{\waj}{\omega_{\mathrm{a}j}}
\newcommand{\Phix}{\Phi_\mathrm{x}}

\newcommand{\ha}{\hat{a}}
\newcommand{\hA}{\hat{a}^\dag}
\newcommand{\hM}{\hat{M}}

\newcommand{\hH}{\hat{H}}

\newcommand{\Gphi}{\Gamma_{\varphi\mathrm{m}}}
\newcommand{\Gmeas}{\Gamma_\mathrm{meas}}

\begin{document}

\title{Fast quantum non-demolition readout from longitudinal qubit-oscillator interaction}

\author{Nicolas Didier}
\affiliation{Department of Physics, McGill University, 3600 rue University, Montreal, Quebec H3A 2T8, Canada}
\affiliation{D\'epartment de Physique, Universit\'e de Sherbrooke, 2500 boulevard de l'Universit\'e, Sherbrooke, Qu\'ebec J1K 2R1, Canada}

\author{J\'er\^{o}me Bourassa}
\affiliation{C\'egep de Granby, 235, rue Saint-Jacques, Granby, Qu\'ebec J2G 9H7}

\author{Alexandre Blais}
\affiliation{D\'epartment de Physique, Universit\'e de Sherbrooke, 2500 boulevard de l'Universit\'e, Sherbrooke, Qu\'ebec J1K 2R1, Canada}
\affiliation{Canadian Institute for Advanced Research, Toronto, Canada}


\begin{abstract}
We show how to realize high-fidelity quantum non-demolition qubit readout using longitudinal qubit-oscillator interaction. This is realized by modulating the longitudinal coupling at the cavity frequency. The qubit-oscillator interaction then acts as a qubit-state dependent drive on the cavity, a situation that is fundamentally different from the standard dispersive case. Single-mode squeezing can be exploited to exponentially increase the signal-to-noise ratio of this readout protocol. We present an implementation of this idea in circuit quantum electrodynamics and a possible multi-qubit architecture.
\end{abstract}

\pacs{
42.50.Dv,	
03.67.-a,	
03.65.Ta,	
42.50.Lc	
}

\maketitle

{\it Introduction --} 
Measurement in quantum information processors is generally realized by entangling a qubit to an ancillary system, with the states of the latter strongly depending on the qubit states. The ability to resolve these ancillary, or pointer, states can then result in a qubit measurement. This readout should be fast, quantum non-demolition (QND) and of high-fidelity. A common approach to realize this is the dispersive regime of cavity QED where the electric dipole moment of an atom strongly couples to the electric field of a high-Q cavity~\cite{haroche:2006a}. In the dispersive regime where the qubit-cavity detuning $\Delta$ is large with respect to the coupling strength $g_\mathrm{x}$, the cavity frequency $\wc$ is modified to take a qubit-state dependent value $\wc\pm\chi$, with $\chi=g_\mathrm{x}^2/\Delta$ the dispersive qubit-cavity interaction. Starting in the vacuum state, a drive then displaces the cavity to qubit-state dependent coherent states $|\alpha_{0,1}\rangle$. Resolving these two pointer states by homodyne detection of the transmitted or reflected signal completes the qubit readout. Combined with recent advances in near quantum-limited amplification~\cite{castellanos-beltran:2008a,bergeal:2010a,eichler:2014a}, this approach has led to high-fidelity readout with superconducting qubits~\cite{vijay:2011a,vijay:2012a,hatridge:2013a,Jeffrey_2014}.

\begin{figure}[h!]
\includegraphics[width=\columnwidth]{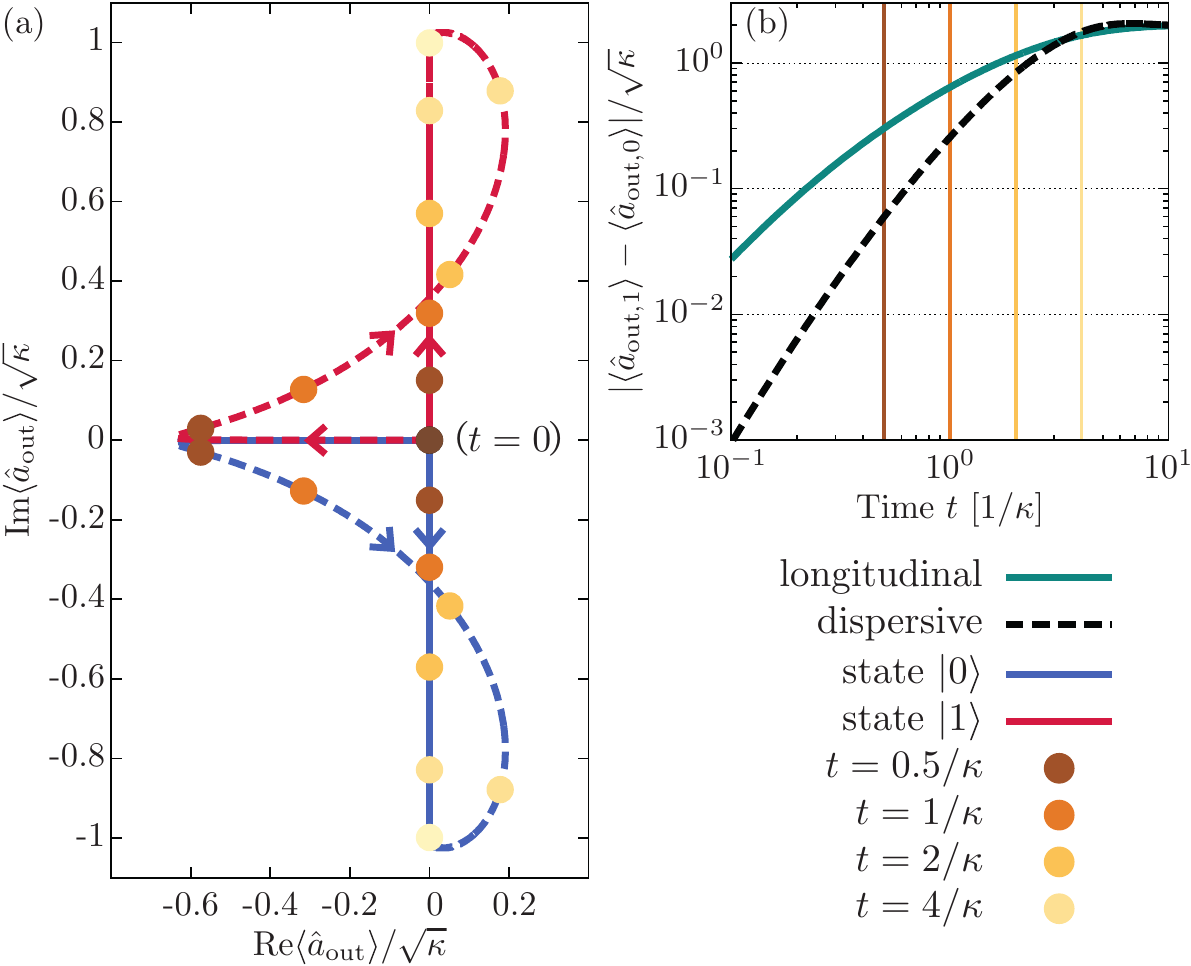}
\caption{(a) Evolution in phase space of the cavity output field $\ha_\out$ for longitudinal (full lines) and dispersive coupling (dashed lines, $\chi=\kappa/2$). Blue and red refer to qubit states. For the chosen parameters, the field starts at $(0,0)$ to reach $(0,\pm1)$ in steady-state; the circles illustrate the position of the pointer states at characteristic times. Longitudinal coupling modulation and coherent drive are switched on as $(2/\pi)\arctan(10\kappa t)$. 
(b) Pointer state separation as a function of time. Vertical lines correspond to the circles of panel (a). 
}
\label{figPhaseSpace}
\end{figure}

Despite these successes, current state-of-the-art dispersive readout suffers from several problems. First, readout must be made faster to meet the stringent requirements of fault-tolerant quantum computation~\cite{raussendorf:2007a}. This is however challenging with the dispersive qubit-cavity interaction taking the form $H_\mathrm{disp} = \chi\hAa\sz$.  Indeed as illustrated by the dashed lines in Fig.~\ref{figPhaseSpace}(a) under this interaction a coherent drive at the input of the cavity first pushes the two pointer states in the same direction in phase space before pulling them apart. There is therefore little information about the qubit state at small measurement times. Second, $H_\mathrm{disp}$ derives in second-order perturbation theory from the electric-dipole interaction $g_\mathrm{x}(\hA+\ha)\sx$~\cite{haroche:2006a}. Since the latter does not commute with the measured qubit observable, $\sz$, dispersive readout is only QND in a perturbative sense. This non-QNDness manifests itself with Purcell decay $\gamma_\kappa = (\gx/\Delta)^2\kappa$~\cite{houck:2008a}, where $\kappa$ is the cavity damping rate, and with the experimentally observed measurement-induced qubit transitions~\cite{Boissonneault_2009,slichter:2012a}. For this reason, the cavity damping rate cannot be made arbitrarily large and the measurement photon number $\bar n$ is typically kept well below the critical photon number $n_\mathrm{crit} = (\Delta/2\gx)^2$~\cite{Blais_2004}. Increasing $\kappa$ and $\bar n$ would otherwise lead to faster qubit measurement.

In this letter, we study an alternative approach that addresses  both the slow measurement time and the non-QNDness. This proposal is based on a longitudinal qubit-cavity interaction of the form $g_\mathrm{z}(\hA+\ha)\sz$. As already noted in Refs.~\cite{kerman:2013a,billangeon:2015a}, this coupling is purely QND and therefore avoids Purcell decay. Here, we show that under appropriate driving, longitudinal interaction leads to an optimal separation of the two pointer states in phase space, without the initial slow separation that is characteristic of the dispersive interaction. In contrast to the dispersive case~\cite{Barzanjeh_2014,didier:2015a}, we moreover find that the signal-to-noise ratio (SNR) of qubit readout can be exponentially improved by injecting a single-mode squeezed state in the cavity. As a possible realization of this idea, we discuss a circuit QED implementation based on a transmon qubit~\cite{Koch_2007} strongly coupled to the flux degree of freedom of an oscillator. The effect of imperfections and a possible multi-qubit architecture are also presented.

{\it Longitudinal readout --}
Under longitudinal coupling, the qubit-cavity Hamiltonian reads ($\hbar=1$)
\begin{equation}
\hat{H}=\wc\hAa +\tfrac{1}{2}\wa\sz+\gz\sz(\hA+\ha),
\end{equation}
where $\wc$ and $\wa$ are respectively the cavity and qubit frequencies, while $\gz$ is the longitudinal coupling strength.  The realization of multi-qubit gates based on this interaction has already been discussed in the context of trapped ions~\cite{milburn:2000a,sorensen:2000a,garcia-ripoll:2003a,leibfried:2003b} and superconducting qubits~\cite{blais:2007a,kerman:2013a,billangeon:2015a}. In particular, Ref.~\cite{kerman:2013a} proposes to modulate the bias of a flux qubit to realize two-qubit gates. In the absence of external perturbations however, this interaction leads in steady-state to a qubit-state dependent displacement of the cavity field of amplitude $\pm\gz/(\wc+i\kappa/2)$. In other words, longitudinal interaction is of no consequences for the typical case where $\wc \gg \gz,\kappa$.

Here we propose to render this interaction resonant for readout by modulating the coupling at the resonator frequency: $\gz(t)=\bgz+\tgz\cos(\wc t)$. In the interaction picture and neglecting fast-oscillating terms we obtain
\begin{equation}
\tilde{H}=\tfrac{1}{2}\tgz\sz(\hA+\ha).
\end{equation}
This now leads to a large qubit-state dependent displacement $\pm\tgz/\kappa$. Even with a conservative modulation amplitude $\tgz\sim 10\kappa$, the steady-state displacement corresponds to 100 photons and the two qubit states are easily distinguishable by homodyne detection. We note that with this longitudinal coupling there is no concept of critical photon number and large intra-cavity photon population is therefore not expected to perturb the qubit.

\begin{figure}
\includegraphics[width=\columnwidth]{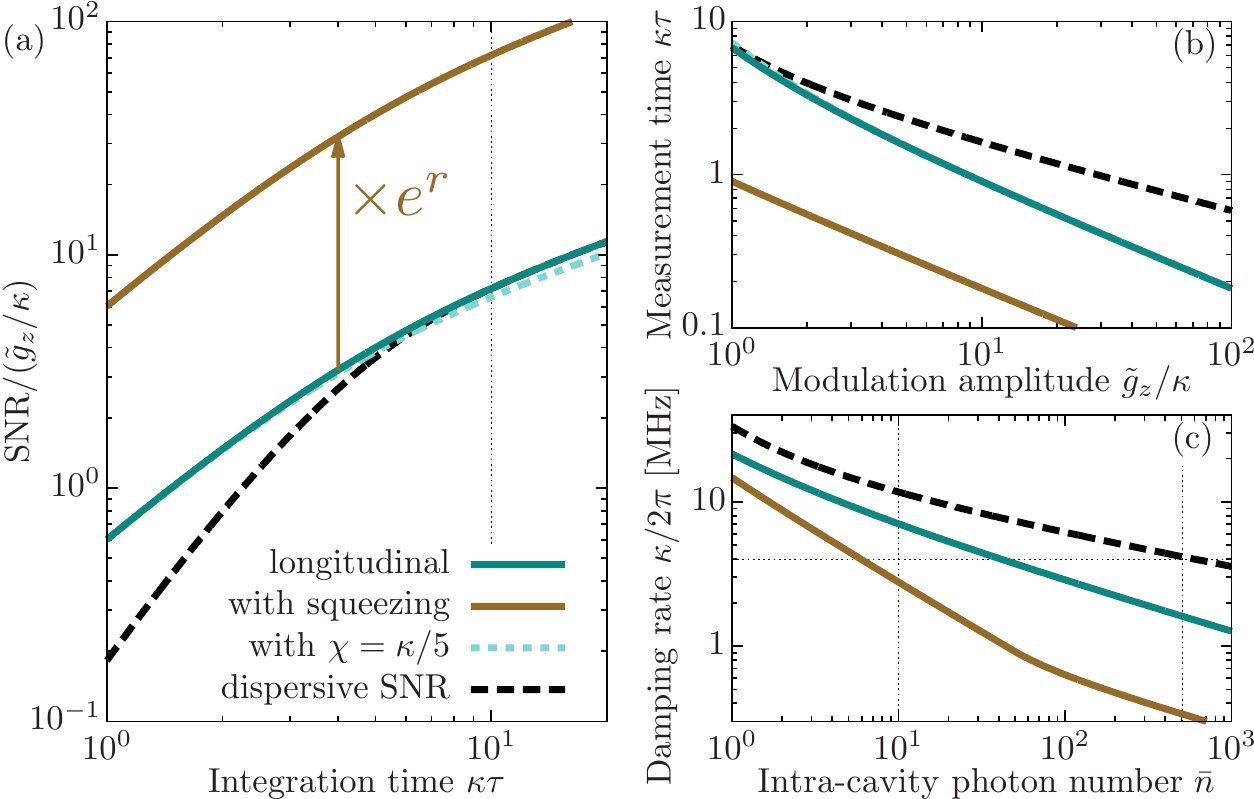}
\caption{
(a) SNR in units of $\tgz/\kappa$ as a function of integration time $\tau$. Longitudinal coupling (green line) is compared to dispersive coupling (dashed black line, $\chi=\kappa/2$) for the same pointer state separation, $|\tgz|=|\epsilon|$.
The dotted cyan line accounts for a residual dispersive coupling of $\chi=\kappa/5$. The full brown line shows the exponential improvement obtained for a single-mode squeezed input state with $\Exp{2r}=20\,\mathrm{dB}$. (b) Integration time $\tau$ required to achieve a fidelity $F=99.99\,\%$ versus longitudinal coupling. (c) Cavity damping rate to reach a fidelity of 99.99\% in 50 ns versus intra-cavity photon number $\bar{n}=(\gz/\kappa)^2=2(\epsilon/\kappa)^2$. Squeezing (full brown line) helps in further reducing the required photon number or cavity decay rate. The squeeze strength is optimized for each $\kappa$, with a maximum set to 20 dB reached close to $\kappa/2\pi = 1$ MHz.
}
\label{figSNR}
\end{figure}

While large pointer state separation can be obtained under a strong measurement tone in the dispersive case, here crucially the separation of the two pointer states occurs much faster. This is illustrated in Fig.~\ref{figPhaseSpace}(a) which shows the path in phase space for both measurement protocols (full lines:~longitudinal; dashed lines:~dispersive). The circles on this figure show the position of the pointer states at characteristic times until steady-state is reached. Clearly, with the proposed scheme the pointer states take the optimal path in phase space towards their maximal separation reached at steady-state. As shown in Fig.~\ref{figPhaseSpace}(b), this leads to a larger pointer state separation at short times.

The consequence of this observation on qubit measurement can be quantified with the signal-to-noise ratio. The SNR is evaluated from $\hM(\tau)=\sqrt{\kappa}\int_0^\tau\mathrm{d}t[\hA_\out(t)+\ha_\out(t)]$, the measurement operator for homodyne detection of the output signal $\ha_\out$ with a measurement time $\tau$. From this expression, the signal is defined as $|\moy\hM_{1}-\moy\hM_{0}|$, where the label $\{0,1\}$ indicates the qubit state, while the imprecision noise is $[\moy{\hM_{N1}^2(\tau)}+\moy{\hM_{N0}^2(\tau)}]^{1/2}$ with $\hM_N=\hM-\moy\hM$~\cite{didier:2015a}. Combining these two expressions, the SNR  for the longitudinal case then reads~\cite{SM}
\begin{equation}\label{eq:SNRgz}
\begin{split}
\SNR_\mathrm{z}&=\sqrt{8}\frac{|\tgz|}{\kappa}\sqrt{\kappa\tau}\left[1-\frac{2}{\kappa\tau}\left(1-\Exp{-\frac{1}{2}\kappa\tau}\right)\right].
\end{split}
\end{equation}
This is to be contrasted to the SNR obtained for dispersive qubit readout under a coherent drive of amplitude $\epsilon$ and optimal dispersive coupling $\chi = \kappa/2$~\cite{didier:2015a,SM,gambetta:2008a}
\begin{equation}\label{eq:SNRdisp}
\begin{split}
\SNR_\chi
&=\sqrt{8}\frac{|\epsilon|}{\kappa}\,\sqrt{\kappa\tau}\left[1-\frac{2}{\kappa\tau}\left(1-\Exp{-\frac{1}{2}\kappa\tau}\cos\tfrac{1}{2}\kappa\tau\right)\right].
\end{split}
\end{equation}
Both expressions have a similar structure, making very clear the similar role of $\tgz$ and $\epsilon$, except for the cosine that is a signature of the complex path in phase space of the dispersive case. Importantly, for short measurement times $\kappa\tau\ll1$, we find a favorable scaling for longitudinal readout with $\SNR_\mathrm{z} \propto \SNR_\chi / \kappa \tau$. This advantage is illustrated in Fig.~\ref{figSNR}(a) that shows the SNR versus integration time for longitudinal (full green line) and dispersive (dashed black line) coupling. At equivalent steady-state separation ($\tgz=\epsilon$), this leads to shorter measurement time for longitudinal coupling. This is made clear in Fig.~\ref{figSNR}(b) presenting the measurement time required to reach a fidelity of $99.99\%$ as a function of the modulation amplitude.

As evidenced by the above discussion, for $\tgz=\epsilon$ the advantage over dispersive readout is found at short integration times. This is especially true when considering the non-perturbative effects that affect the QNDness of dispersive readout.  As an  illustration of this, Fig.~\ref{figSNR}(c) shows the cavity damping rate vs photon number required to reach a fidelity of 99.99\% in the short measurement time $\tau = 50$ ns. The full green line again represents longitudinal readout and the dashed black line the dispersive case. The horizontal dotted line at $\kappa/2\pi = 4$ MHz is a typical value for circuit QED experiments and in particular corresponds to Ref.~\cite{Jeffrey_2014} where a fidelity of 99.8\% was achieved in 140 ns. With this $\kappa$, reaching a fidelity of 99.99\% in 50 ns with dispersive readout would require as many as 500 photons (see vertical dotted line). For most circuit QED experiments, this is well above $n_\mathrm{crit}$ where non-perturbative effects are expected to reduce the readout fidelity. On the other hand, reaching the same goal with longitudinal readout only requires $\sim 40$ photons. Importantly, working at larger photon number is also a possibility here since there is no critical photon number. Alternatively, working with $\sim 10$ photons requires a large cavity damping rate $\kappa/2\pi \sim 10$ MHz to reach the above goal. Under transverse coupling, this would lead to significant Purcell decay. Indeed, for the typical value $\gx/\Delta \sim 1/10$, the cavity-induced relaxation time $1/\gamma_\kappa \sim 1.6~\mu$s is much smaller than current qubit relaxation times. As already mentioned above, longitudinal coupling does not lead to Purcell decay~\cite{kerman:2013a,billangeon:2015a}.

In short, the proposed approach allows to reach large readout fidelities in short measurement times. Reaching the same goal with dispersive readout requires either large $\kappa$ or large $\bar n$, something that in practice would lead to a reduction of the readout fidelity.
It is also interesting to point out that longitudinal readout saturates the inequality $\Gphi \ge \Gmeas$ linking the measurement-induced dephasing rate $\Gphi$ to the measurement rate $\Gmeas$ and is therefore quantum limited~\cite{SM}. 

{\it Single-mode squeezing --}
The SNR of longitudinal readout can also be exponentially improved with a single-mode squeezed input state on the cavity. For this it suffices to chose the squeeze axis to be orthogonal to the qubit-state dependent displacement generated by $\gz(t)$. In Fig.~\ref{figPhaseSpace}(a), this corresponds to orienting the squeeze axis along the vertical axis. With this choice, and since the squeeze angle is unchanged under evolution with longitudinal coupling, the imprecision noise is exponentially reduced and the signal-to-noise ratio simply becomes $e^r \SNR_\mathrm{z}$, with $r$ the squeeze parameter~\cite{SM}. This exponential enhancement is apparent from the full brown line in Fig.~\ref{figSNR}(a) and in the corresponding reduction of the measurement time in Fig.~\ref{figSNR}(b). 


This is in stark contrast to standard dispersive readout where single-mode squeezing can lead to an \emph{increase} of the measurement time~\cite{Barzanjeh_2014,didier:2015a}. Indeed, under dispersive coupling, the squeeze angle undergoes a qubit-state dependent rotation. As a result, both the squeezed and the anti-squeezed quadrature contributes to the imprecision noise. We note that the situation can be different in the presence of two-mode squeezing~\cite{Barzanjeh_2014} where the present exponential increase in SNR can be recovered by engineering the dispersive coupling of the qubit to two cavities~\cite{didier:2015a}.

\begin{figure}[t]
\includegraphics[width=1\columnwidth]{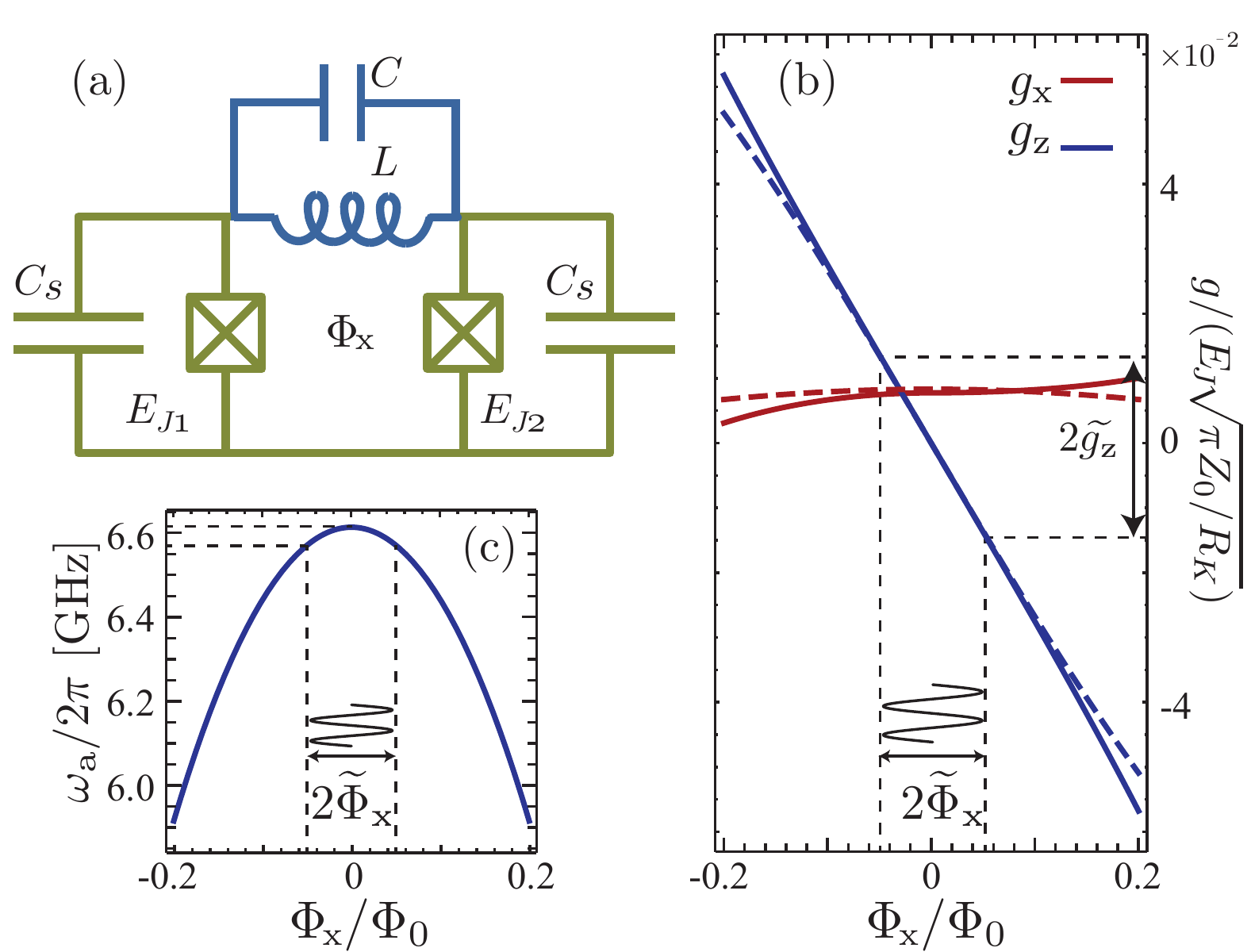}
\caption{(a) Circuit QED implementation of longitudinal coupling with a transmon qubit of Josephson energies $E_{J1} = E_J(1+d)/2$, $E_{J2} = E_J(1-d)/2$ with $d\in[0,1]$. (b) $\gz$ and $\gx$ versus flux. Around $\Phix = 0$, $\gz$ as a linear dependence with flux. The spurious transverse coupling $\gx$ results from qubit asymmetry, here $d=0.02$. The dashed lines represent the asymptotic expressions Eqs.~\eqref{eq:gz} and~\eqref{eq:gx}, the full lines  to exact numerical result~\cite{SM}. (c) Transmon frequency versus flux for $E_J/h = 20$~GHz, $E_J/E_C = 67$ and $d=0.02$.
}
\label{figcircuit}
\end{figure}

{\it Circuit QED implementation -- } We now turn to a possible realization of this protocol in circuit QED. Longitudinal coupling of a flux or a transmon qubit to a LC oscillator was already discussed in Refs.~\cite{kerman:2013a,billangeon:2015a}. There, longitudinal coupling results from the mutual inductive coupling between a flux-tunable qubit and the oscillator.  As another example, we follow the general approach developed in Ref.~\cite{Bourassa_2009} and focus on a transmon qubit that is phase-biased by the oscillator. Fig.~\ref{figcircuit}(a) schematically represents a lumped version of this circuit. In practice, the inductors can be replaced by a junction array~\cite{manucharyan:2009b}, both to increase the coupling and to reduce the size of the qubit's flux-bias loop. An in-depth analysis of an alternative realization based on a transmission-line resonator can be found in Ref.~\cite{SM}.

The Hamiltonian of the circuit of Fig.~\ref{figcircuit}(a) is similar to that of a flux-tunable transmon, but where the external flux $\Phi_\mathrm{x}$ is replaced by $\Phi_\mathrm{x} + \delta$ with $\delta$ the phase drop at the oscillator~\cite{vion:2002a}. Taking the junction capacitances to be equal and assuming for simplicity that $Z_0/R_K\ll 1$ with $Z_0 = \sqrt{L/C}$ and $R_K$ the resistance quantum, this Hamiltonian can be expressed as $\hH = \hH_\mathrm{r} + \hat H_\mathrm{q} + \hH_\mathrm{qr}$. In this expression, $\hH_\mathrm{r} = \wc \hAa$ is the oscillator Hamiltonian and $\hH_\mathrm{q} = \wa\sz/2$ is the Hamiltonian of a flux-tunable transmon that we write here in its two-level approximation~\cite{Koch_2007}. The qubit-oscillator interaction takes the form $\hH_\mathrm{qr} = \gx(\hA+\ha)\sx+g_z(\hA+\ha)\sz$ with~\cite{SM}
\begin{align}
\label{eq:gz}
\gz &= -\frac{E_J}{2}\left(\frac{2E_C}{E_J}\right)^{1/2}\sqrt{\frac{\pi Z_0}{R_K}}\sin\left(\frac{\pi\Phix}{\Phi_0}\right),
\\
\label{eq:gx}
\gx & = d E_J\left(\frac{2E_C}{E_J}\right)^{1/4}\sqrt{\frac{\pi Z_0}{R_K}}\cos\left(\frac{\pi\Phix}{\Phi_0}\right),
\end{align} 
and where $E_J$ is the mean Josephson energy, $d$ the Josephson energy asymmetry and $E_C$ the qubit's charging energy. Expressions for these quantities in terms of the elementary circuit parameters are given in Ref.~\cite{SM}. As desired, the transverse coupling $\gx$ vanishes exactly for a symmetric transmon with $d=0$, leaving only longitudinal coupling $\gz$. Thanks to the phase bias rather than inductive coupling, $\gz$ can be made large~\cite{Bourassa_2009}. For example, with the realistic values $E_J/h = 20$ GHz, $E_J/E_C = 67$ and $Z_0 = 50~\Omega$ we find $\gz/2\pi \approx  135~\mathrm{MHz} \times \sin\left(\pi\Phix/\Phi_0\right)$. The flux dependence of both $\gz$ (blue lines) and $\gx$ (red lines) for a finite asymmetry $d=0.02$ are illustrated in Fig.~\ref{figcircuit}(b). The dashed lines refer to the above asymptotic expressions for $\gz$ and $\gx$ while the full lines are exact numerical results. Modulating the flux by $0.05\Phi_0$ around $\Phix =0$, we find $\tgz/2\pi \sim 21$~MHz. This is accompanied by a small change of the qubit frequency of $\sim 40$ MHz, see Fig.~\ref{figcircuit}(c). Importantly, this frequency change does not affect the SNR under longitudinal readout~\cite{SM}. 



{\it Tolerance to imperfections -- } 
When $d\neq0$, a finite transversal coupling is present. This is illustrated in Fig.~\ref{figcircuit}(b) where for a realistic value of $d=0.02$~\cite{fink:2009b} and the above parameters we find $\gx/2\pi \approx  13 ~\mathrm{MHz} \times \cos\left(\pi\Phix/\Phi_0\right)$. The effect of this unwanted coupling can be mitigated by working at large qubit-resonator detuning $\Delta$ where the resulting dispersive interaction $\chi = \gx^2/\Delta$ can be made small. For exemple, the above numbers correspond to a detuning of $\Delta/2\pi = 3$~GHz where $\chi/2\pi \sim 5.6$ kHz. It is important to emphasize that, contrary to dispersive readout, the longitudinal approach is not negatively affected by a large detuning. The large detuning moreover reduces Purcell decay which, for the above $\gx$, can very easily be made small enough to avoid the qubit from being Purcell limited.

When considering higher-order terms in $Z_0/R_K$, the Hamiltonian of the circuit of Fig.~\ref{figcircuit}(a) contains a dispersive-like interaction $\chiz\hAa\sz$ even at $d=0$. For the parameters already used above, we find $\chiz/2\pi \sim$ 5.3 MHz~\cite{SM}. Contrary to the standard dispersive coupling, $\chiz$ cannot be made small by detuning the qubit from the resonator. However since it is not derived from a transverse coupling, it is not linked to any Purcell decay. Moreover, at small integration times $\SNR_\mathrm{z}$ is not affected by a finite dispersive-like coupling~\cite{SM}. This is illustrated in Fig.~\ref{figSNR}(a) where the cyan dotted line, corresponding $\SNR_\mathrm{z}$ in the presence of $\chi=\kappa/5$, is barely distinguishable from the ideal case. 


Finally, when discussing the exponential gain in SNR provided by single-mode squeezing, we have assumed a source of broadband pure squeezing. The effect of a finite squeezing bandwidth $\Gamma$ was already studied in Ref.~\cite{didier:2015a} and only leads to a small reduction of the SNR for $\Gamma \gg \kappa$. On the other hand, deviation from unity of the squeezing purity $P$ leads to a reduction of the SNR by $1/\sqrt{P}$.
The SNR being decoupled from the anti-squeezed quadrature, the purity simply renormalizes the squeeze parameter.

{\it Multi-qubit architecture -- }
A possible mutli-qubit architecture consists of qubits longitudinally coupled to a readout resonator (of annihilation operator $\ha_\mathrm{z}$) and transversally coupled to a high-Q bus resonator ($\ha_\mathrm{x}$). 
The Hamiltonian describing this system is 
\begin{equation}
\begin{split}
\hat{H} 
&=
\omega_{\mathrm{rz}} \hat a_\mathrm{z}^\dag \hat a_\mathrm{z} + \omega_{\mathrm{rx}} \hat a_\mathrm{x}^\dag \hat a_\mathrm{x} + \sum_j \tfrac{1}{2}\waj\szj\\
& +\sum_j\gzj\szj(\hat a_\mathrm{z}^\dag+\hat a_\mathrm{z}) +\sum_j\gxj\sxj(\hat a_\mathrm{x}^\dag+\hat a_\mathrm{x}).
\end{split}
\end{equation}
In this architecture, readout is realized by taking advantage of the longitudinal coupling while logical operation are realized using the bus resonator. A scalable architecture taking advantage of longitudinal coupling is discussed at length in Ref.~\cite{billangeon:2015a}. Here, we again consider that the longitudinal coupling of each qubit can be modulated independently. We take this modulation to be $\gzj(t)=\bgz+\tgz\cos(\wc t+\varphi_j)$ where the phase $\varphi_j$ is adjustable. In the interaction picture and neglecting fast-oscillating terms, the longitudinal coupling becomes 
\begin{equation}
\tilde{H}_\mathrm{z}= \Big(\tfrac{1}{2}\tgz\sum_j\szj\Exp{-i\varphi_j}\Big) \hat a_\mathrm{z} + \mathrm{h.c.}
\end{equation}
This effective drive on the resonator displaces the field to a multi-qubit-state dependent coherent state allowing single-shot multi-qubit measurements. 
For two qubits, $\varphi_j=j\pi/2$ leads to four well separated states in phase space. 
Other choices of phase lead to overlapping pointer states corresponding to different multi-qubit states. 
Exemples are $\varphi_j=0$ for which the two-qubit states $\ket{01}$ and $\ket{10}$ are indistinguishable, and
 $\varphi_j=j\pi$ where these states are replaced by $\ket{00}$ and $\ket{11}$. This can be exploited to create entanglement by measurement~\cite{lalumiere:2010a}. The 3-qubit GHZ state is obtained with $\varphi_j=j2\pi/3$~\cite{SM}.


{\it Conclusion-- }
We have shown that modulating longitudinal coupling between a qubit and an oscillator leads to fast QND qubit readout. Because of the optimal motion of the pointer states in phase, the measurement time is reduced with respect to standard dispersive readout.  For the same reason, this approach can be further improved by using single-mode squeezing. 

{\it Acknowledgements-- }
We thank A.~Clerk for useful discussions. This work was supported by the Army Research Office under Grant W911NF-14-1-0078, INTRIQ and NSERC.


%

\clearpage


\section*{Supplementary material}

# Supplemental Material for
# "Fast quantum non-demolition readout from longitudinal qubit-oscillator interaction"

Nicolas Didier,[1,2] Jérôme Bourassa,[3] and Alexandre Blais[2,4]

[1]*Department of Physics, McGill University, 3600 rue University, Montreal, Quebec H3A 2T8, Canada*
[2]*Départment de Physique, Université de Sherbrooke,*
*2500 boulevard de l'Université, Sherbrooke, Québec J1K 2R1, Canada*
[3]*Cégep de Granby, 235, rue Saint-Jacques, Granby, Québec J2G 9H7*
[4]*Canadian Institute for Advanced Research, Toronto, Canada*

## I. SIGNAL-TO-NOISE RATIO

We derive the signal-to-noise ratio (SNR) for different qubit readout protocols: Longitudinal coupling in Sec. I A; Longitudinal coupling with single-mode squeezed states in Sec. I B; Standard dispersive coupling in Sec. I C; Longitudinal and transverse couplings in the dispersive regime in Sec. I D.

### A. Longitudinal coupling

#### 1. Modulation at the resonator frequency

We consider a qubit longitudinally coupled to a resonator with the Hamiltonian

$$\hat{H} = \omega_\mathrm{r} \hat{a}^\dagger \hat{a} + \tfrac{1}{2}\omega_\mathrm{a} \hat{\sigma}_z + [g_\mathrm{z}(t)\hat{a}^\dagger + g_\mathrm{z}^*(t)\hat{a}]\hat{\sigma}_z. \tag{S1}$$

In this expression, $\omega_\mathrm{r}$ is the resonator frequency, $\omega_\mathrm{a}$ the qubit frequency and $g_\mathrm{z}$ the longitudinal coupling that is modulated at the resonator frequency:

$$g_\mathrm{z}(t) = \bar{g}_\mathrm{z} + |\tilde{g}_\mathrm{z}|\cos(\omega_\mathrm{r} t + \varphi). \tag{S2}$$

In the interaction picture and using the rotating-wave approximation (RWA), the above Hamiltonian simplifies to

$$\hat{H} = \tfrac{1}{2}[\tilde{g}_\mathrm{z}\hat{a}^\dagger + \tilde{g}_\mathrm{z}^*\hat{a}]\hat{\sigma}_z, \tag{S3}$$

where $\tilde{g}_\mathrm{z} \equiv |\tilde{g}_\mathrm{z}|e^{i\varphi}$ is the modulation amplitude. From Eq. (S3), it is clear that the modulated longitudinal coupling plays the role of a qubit-state dependent drive. The Langevin equation of the cavity field simply reads

$$\dot{\hat{a}} = -i\tfrac{1}{2}\tilde{g}_\mathrm{z}\hat{\sigma}_z - \tfrac{1}{2}\kappa\hat{a} - \sqrt{\kappa}\hat{a}_\mathrm{in}, \tag{S4}$$

where $\hat{a}_\mathrm{in}$ is the input field [S1]. Taking this input to be the vacuum, the input correlations are then defined by $\langle \hat{a}_\mathrm{in}(t)\hat{a}_\mathrm{in}^\dagger(t')\rangle = [\hat{a}_\mathrm{in}(t), \hat{a}_\mathrm{in}^\dagger(t')] = \delta(t-t')$. Using the input-output boundary condition $\hat{a}_\mathrm{out} = \hat{a}_\mathrm{in} + \sqrt{\kappa}\hat{a}$ [S1], integration of the Langevin equation leads to

$$\alpha_\mathrm{out}(t) = -\frac{i\tilde{g}_\mathrm{z}}{\sqrt{\kappa}}\langle\hat{\sigma}_z\rangle\left[1 - e^{-\tfrac{1}{2}\kappa t}\right], \qquad \hat{d}_\mathrm{out}(t) = \hat{d}_\mathrm{in}(t) - \kappa\int_{-\infty}^{t}\mathrm{d}t' e^{-\tfrac{1}{2}\kappa(t-t')}\hat{d}_\mathrm{in}(t'), \tag{S5}$$

where $\alpha_\mathrm{out} = \langle\hat{a}_\mathrm{out}\rangle$ stands for the output field mean value and $\hat{d}_\mathrm{out} = \hat{a}_\mathrm{out} - \alpha_\mathrm{out}$ its fluctuations. Because here the qubit-dependent drive comes from modulations of the coupling, and not from an external coherent drive, there is no interference between the outgoing and the input fields. As a result, $\alpha = \alpha_\mathrm{out}/\sqrt{\kappa}$ and the intracavity photon number evolves as

$$\langle\hat{a}^\dagger\hat{a}\rangle = \frac{|\tilde{g}_\mathrm{z}|^2}{\kappa^2}\left[1 - e^{-\tfrac{1}{2}\kappa t}\right]^2. \tag{S6}$$

The measurement operator corresponding to homodyne detection of the output signal with an integration time $\tau$ and homodyne angle $\phi_\mathrm{h}$ is

$$\hat{M}(\tau) = \sqrt{\kappa}\int_0^\tau \mathrm{d}t[\hat{a}_\mathrm{out}^\dagger(t)e^{i\phi_\mathrm{h}} + \hat{a}_\mathrm{out}(t)e^{-i\phi_\mathrm{h}}]. \tag{S7}$$

The signal for such a measurement is $\langle \hat{M} \rangle$ while the noise operator is $\hat{M}_N = \langle \hat{M} \rangle$. In the presence of a qubit, the measurement signal is then

$$\langle \hat{M} \rangle_1 - \langle \hat{M} \rangle_0 = 4|\tilde{g}_z| \sin(\varphi - \phi_h)\, \tau \left[1 - \frac{2}{\kappa\tau}\left(1 - e^{-\frac{1}{2}\kappa\tau}\right)\right]. \tag{S8}$$

On the other hand, the measurement noise is equal to $\langle \hat{M}_N^2(\tau) \rangle = \kappa\tau$. Combining these two expressions, the signal-to-noise ratio (SNR) then reads

$$\mathrm{SNR}^2 \equiv \frac{|\langle \hat{M} \rangle_1 - \langle \hat{M} \rangle_0|^2}{\langle \hat{M}_{N1}^2(\tau) \rangle + \langle \hat{M}_{N0}^2(\tau) \rangle} \tag{S9}$$

$$= \frac{8|\tilde{g}_z|^2}{\kappa^2} \sin^2(\varphi - \phi_h)\, \kappa\tau \left[1 - \frac{2}{\kappa\tau}\left(1 - e^{-\frac{1}{2}\kappa\tau}\right)\right]^2. \tag{S10}$$

The SNR is optimized by choosing the modulation phase $\varphi$ and the homodyne angle such that $\varphi - \phi_h = \frac{\pi}{2} \mod \pi$. With this choice, the optimized SNR finally reads

$$\mathrm{SNR} = \sqrt{8}\frac{|\tilde{g}_z|}{\kappa} \sqrt{\kappa\tau} \left[1 - \frac{2}{\kappa\tau}\left(1 - e^{-\frac{1}{2}\kappa\tau}\right)\right]. \tag{S11}$$

At long measurement times ($\tau \gg 1/\kappa$), the signal-to-noise ratio evolves as $\mathrm{SNR} = \sqrt{8}\frac{|\tilde{g}_z|}{\kappa}\sqrt{\kappa\tau}$ while in the more experimentally interesting case of short measurement times we find $\mathrm{SNR} = \frac{1}{\sqrt{2}}\frac{|\tilde{g}_z|}{\kappa}(\kappa\tau)^{3/2}$. In short, the SNR increases as $\tau^{3/2}$, much faster than in the dispersive regime where the SNR rather increases as $\tau^{5/2}$ (cf. Sec. I C).

### 2. Measurement and dephasing rates

Following Ref. [S2], to evaluate the measurement-induced dephasing rate we apply a polaron-type transformation on Hamiltonian Eq. (S3) consisting of a displacement of $\hat{a}$ by $-i\tilde{g}_z\hat{\sigma}_z/\kappa$. Under this transformation, the cavity decay Lindbladian $\kappa\mathcal{D}[\hat{a}]\hat{\rho} = \mathcal{D}[\hat{a}]\rho = \hat{a}\rho\hat{a}^\dagger - \frac{1}{2}\{\hat{a}^\dagger\hat{a}, \rho\}$, leads to $\frac{1}{2}\Gamma_{\varphi m}\mathcal{D}[\hat{\sigma}_z]\hat{\rho}$ where $\Gamma_{\varphi m} = 2|\tilde{g}_z|^2/\kappa$ is the measurement-induced dephasing. On the other hand, the measurement rate is obtained from the SNR as $\Gamma_{\mathrm{meas}} = \mathrm{SNR}^2/(4\tau) = 2|\tilde{g}_z|^2/\kappa$ [S3]. The relation between the dephasing and the measurement rate is then $\Gamma_{\mathrm{meas}} = \Gamma_{\varphi m}$. This is the bound reached for a quantum limited measurement [S3].

### 3. Modulation bandwidth

We now consider the situation where the longitudinal coupling is modulated at a frequency $\omega_m \neq \omega_r$, i.e. $g_z(t) = \bar{g}_z + |\tilde{g}_z|\cos(\omega_m t + \varphi)$. Assuming that the detuning $\Delta_m = \omega_m - \omega_r$ is small with respect to the modulation amplitude $\tilde{g}_z$, the Hamiltonian in a frame rotating at the modulation frequency now reads under the RWA as

$$\hat{H} = -\Delta_m \hat{a}^\dagger \hat{a} + \tfrac{1}{2}[\tilde{g}_z \hat{a}^\dagger + \tilde{g}_z^* \hat{a}]\hat{\sigma}_z, \tag{S12}$$

The corresponding Langevin equation is then

$$\dot{\hat{a}} = -i\tfrac{1}{2}\tilde{g}_z\hat{\sigma}_z + i\Delta_m \hat{a} - \tfrac{1}{2}\kappa\hat{a} - \sqrt{\kappa}\hat{a}_{\mathrm{in}}, \tag{S13}$$

yielding for the output field,

$$\alpha_{\mathrm{out}}(t) = -\frac{i\tilde{g}_z\sqrt{\kappa}}{\kappa - 2i\Delta_m}\langle \hat{\sigma}_z \rangle \left[1 - e^{(i\Delta_m - \frac{1}{2}\kappa)t}\right], \qquad \hat{d}_{\mathrm{out}}(t) = \hat{d}_{\mathrm{in}}(t) - \kappa \int_{-\infty}^t \mathrm{d}t'\, e^{(i\Delta_m - \frac{1}{2}\kappa)(t-t')} \hat{d}_{\mathrm{in}}(t'). \tag{S14}$$

From these expressions, the measurement signal is then

$$\langle \hat{M} \rangle_1 - \langle \hat{M} \rangle_0 = \frac{4|\tilde{g}_z|}{\sqrt{1 + \left(\frac{2\Delta_m}{\kappa}\right)^2}} \sin[\varphi - \phi_h + \arctan(2\Delta_m/\kappa)]\,\tau$$

$$\qquad - \frac{8|\tilde{g}_z|/\kappa}{1 + \left(\frac{2\Delta_m}{\kappa}\right)^2} \sin[\varphi - \phi_h + 2\arctan(2\Delta_m/\kappa)]$$

$$\qquad + \frac{8|\tilde{g}_z|/\kappa}{1 + \left(\frac{2\Delta_m}{\kappa}\right)^2} \sin[\varphi - \phi_h + 2\arctan(2\Delta_m/\kappa) + \Delta_m\tau]\, e^{-\frac{1}{2}\kappa\tau}. \tag{S15}$$

While the signal is changed, the noise is however not modified by the detuning. From the above expression, given a detuning $\Delta_m$ and a measurement time $\tau$, there is an optimal angle $\varphi$ that maximizes the SNR.

### B. Longitudinal coupling with squeezing

We now consider the situation where the modulation detuning is zero and where the input field is in a single-mode squeezed vacuum. This leaves the signal unchanged, but has we now show leads to an exponential increase of the SNR with the squeeze parameter $r$. Indeed, in the frame of the resonator, the correlations of the bath fluctuations are now

$$\begin{pmatrix} \langle \hat{d}_{\text{in}}^\dagger(t)\hat{d}_{\text{in}}(t')\rangle & \langle \hat{d}_{\text{in}}(t)\hat{d}_{\text{in}}(t')\rangle \\ \langle \hat{d}_{\text{in}}^\dagger(t)\hat{d}_{\text{in}}^\dagger(t')\rangle & \langle \hat{d}_{\text{in}}(t)\hat{d}_{\text{in}}^\dagger(t')\rangle \end{pmatrix} = \begin{pmatrix} \sinh^2 r & \frac{1}{2}\sinh 2r\, e^{2i\theta} \\ \frac{1}{2}\sinh 2r\, e^{-2i\theta} & \cosh^2 r \end{pmatrix}\delta(t-t'). \tag{S16}$$

where we have assumed broadband squeezing with a squeeze angle $\theta$. The measurement noise is then

$$\langle \hat{M}_N^2(\tau)\rangle = \kappa \int_0^\tau dt \int_0^\tau dt' \Big[\langle \hat{d}_{\text{out}}^\dagger(t)\hat{d}_{\text{out}}(t')\rangle + \langle \hat{d}_{\text{out}}(t)\hat{d}_{\text{out}}^\dagger(t')\rangle \\ + \langle \hat{d}_{\text{out}}(t)\hat{d}_{\text{out}}(t')\rangle e^{-2i\phi_\text{h}} + \langle \hat{d}_{\text{out}}^\dagger(t)\hat{d}_{\text{out}}^\dagger(t')\rangle e^{2i\phi_\text{h}}\Big]. \tag{S17}$$

The output-field correlations are easily obtained from

$$\begin{pmatrix} \langle \hat{d}_{\text{out}}^\dagger(t)\hat{d}_{\text{out}}(t')\rangle & \langle \hat{d}_{\text{out}}(t)\hat{d}_{\text{out}}(t')\rangle \\ \langle \hat{d}_{\text{out}}^\dagger(t)\hat{d}_{\text{out}}^\dagger(t')\rangle & \langle \hat{d}_{\text{out}}(t)\hat{d}_{\text{out}}^\dagger(t')\rangle \end{pmatrix} = \begin{pmatrix} \langle \hat{d}_{\text{in}}^\dagger(t)\hat{d}_{\text{in}}(t')\rangle & \langle \hat{d}_{\text{in}}(t)\hat{d}_{\text{in}}(t')\rangle \\ \langle \hat{d}_{\text{in}}^\dagger(t)\hat{d}_{\text{in}}^\dagger(t')\rangle & \langle \hat{d}_{\text{in}}(t)\hat{d}_{\text{in}}^\dagger(t')\rangle \end{pmatrix} \tag{S18}$$

which holds here since the drive is 'internal' to the cavity. As a result

$$\langle \hat{M}_N^2(\tau)\rangle = \{\cosh(2r) + \sinh(2r)\cos[2(\phi_\text{h} - \theta)]\}\kappa\tau. \tag{S19}$$

The noise is minimized by choosing $\theta$ according to $\theta - \phi_\text{h} = \frac{\pi}{2} \mod \pi$. With this choice, the SNR reads

$$\text{SNR}(r) = e^r \text{SNR}(r=0). \tag{S20}$$

The SNR is thus exponentially enhanced, leading to Heisenberg-limited scaling as discussed in Ref. [S4].

### C. Dispersive coupling

For completeness, we derive here the SNR for dispersive readout; this result can already be found in Ref. [S2]. In the dispersive regime, the qubit-cavity Hamiltonian reads

$$\hat{H} = \omega_\text{r}\hat{a}^\dagger\hat{a} + \tfrac{1}{2}\omega_\text{a}\hat{\sigma}_z + \chi\hat{a}^\dagger\hat{a}\hat{\sigma}_z, \tag{S21}$$

where $\chi = g_\text{x}^2/\Delta$ is the dispersive shift. The Langevin equation of the cavity field in the interaction picture then reads

$$\dot{\hat{a}} = -i\chi\hat{\sigma}_z\hat{a} - \tfrac{1}{2}\kappa\hat{a} - \sqrt{\kappa}\hat{a}_\text{in}. \tag{S22}$$

With a drive of amplitude $\epsilon = |\epsilon|e^{i\varphi_d}$ on the cavity at resonance, the input field is defined by its mean $\alpha_\text{in} = \langle \hat{a}_\text{in}\rangle = -\epsilon/\sqrt{\kappa}$ and fluctuations $\hat{d}_\text{in} = \hat{a}_\text{in} - \alpha_\text{in}$. Integrating the Langevin equation yields

$$\alpha_\text{out}(t) = \frac{\epsilon}{\sqrt{\kappa}} e^{-i\varphi_\text{qb}\langle\hat{\sigma}_z\rangle}\left[1 - 2e^{-(i\chi\langle\hat{\sigma}_z\rangle + \frac{1}{2}\kappa)t + \frac{1}{2}i\varphi_\text{qb}\langle\hat{\sigma}_z\rangle}\right], \tag{S23}$$

$$\hat{d}_\text{out}(t) = \hat{d}_\text{in}(t) - \kappa\int_{-\infty}^t dt'\, e^{-(i\chi\langle\hat{\sigma}_z\rangle + \frac{1}{2}\kappa)(t-t')}\hat{d}_\text{in}(t'), \tag{S24}$$

where $\varphi_\text{qb} = 2\arctan(2\chi/\kappa)$ is the qubit-induced phase of the output field. Moreover, the intracavity photon number is as

$$\langle \hat{a}^\dagger\hat{a}\rangle = \left(\frac{2|\epsilon|}{\kappa}\right)^2 \cos^2(\tfrac{1}{2}\varphi_\text{qb})\left[1 - 2\cos(\chi t)e^{-\frac{1}{2}\kappa t} + e^{-\kappa t}\right]. \tag{S25}$$

From the above expressions, the measurement signal is

$$M_{S,|1\rangle} - M_{S,|0\rangle} = 4|\epsilon|\sin(\varphi_{\rm qb})\sin(\varphi_d - \phi_{\rm h})\tau\left\{1 - \frac{4}{\kappa\tau}\left[\cos^2(\tfrac{1}{2}\varphi_{\rm qb}) - \frac{\sin(\chi\tau+\varphi_{\rm qb})}{\sin(\varphi_{\rm qb})}e^{-\frac{1}{2}\kappa\tau}\right]\right\}. \quad (S26)$$

On the other hand, the measurement noise is simply equal to $\langle \hat{M}_N^2(\tau)\rangle = \kappa\tau$. The measurement signal is optimized for $\varphi_d - \phi_{\rm h} = \frac{\pi}{2}$ mod $\pi$ and at long integration times by $\varphi_{\rm qb} = \frac{\pi}{2}$, or equivalently $\chi = \kappa/2$ [S2]. For this optimal choice, the SNR then reads

$$\text{SNR} = \sqrt{8}\frac{|\epsilon|}{\kappa}\sqrt{\kappa\tau}\left[1 - \frac{2}{\kappa\tau}\left(1 - e^{-\frac{1}{2}\kappa\tau}\cos\tfrac{1}{2}\kappa\tau\right)\right]. \quad (S27)$$

At long measurement times, the SNR evolves as $\text{SNR} = \sqrt{8}\frac{|\epsilon|}{\kappa}\sqrt{\kappa\tau}$ and at short measurement times it starts as $\text{SNR} = \frac{1}{\sqrt{18}}\frac{|\epsilon|}{\kappa}(\kappa\tau)^{5/2}$.

### D.  Effect of a residual transverse coupling

We consider the presence of a spurious transverse coupling $g_{\rm x}$ in addition to the longitudinal coupling $g_{\rm z}$,

$$\hat{H} = \omega_{\rm r}\hat{a}^\dagger\hat{a} + \omega_{\rm a}\hat{\sigma}_z + \{[\bar{g}_{\rm x} + \tilde{g}_{\rm x}\cos(\omega_{\rm r}t + \varphi_{\rm x})]\hat{\sigma}_x + [\bar{g}_{\rm z} + \tilde{g}_{\rm z}\cos(\omega_{\rm r}t + \varphi)]\hat{\sigma}_z\}(\hat{a}^\dagger + \hat{a}). \quad (S28)$$

We now assume that $g_{\rm x} \ll \Delta$ and we follow the standard approach to eliminate the transverse coupling [S5]. To leading order in $g_{\rm x}/\Delta$ and under the RWA, we find in the interaction picture

$$\tilde{H} = \tfrac{1}{2}(\chi_{\rm x} - 2\chi_{\rm xz})\hat{\sigma}_z + \chi\hat{a}^\dagger\hat{a}\hat{\sigma}_z + \tfrac{1}{2}[\tilde{g}_{\rm z}\hat{a}^\dagger + \tilde{g}_{\rm z}^*\hat{a}], \quad (S29)$$

with the dispersive shifts $\chi = \chi_{\rm x} - 4\chi_{\rm xz}$, $\chi_{\rm x} = \bar{g}_{\rm x}^2/\Delta$ and $\chi_{\rm xz} = \bar{g}_{\rm x}\bar{g}_{\rm z}/\Delta$.

Going to an interaction picture also with respect to the first term of Eq. (S29), our starting point is

$$\hat{H} = \chi\hat{a}^\dagger\hat{a}\hat{\sigma}_z + \tfrac{1}{2}[\tilde{g}_{\rm z}\hat{a}^\dagger + \tilde{g}_{\rm z}^*\hat{a}]\hat{\sigma}_z. \quad (S30)$$

This leads to the Langevin equation

$$\dot{\hat{a}} = -i\tfrac{1}{2}\tilde{g}_{\rm z}\hat{\sigma}_z - (i\chi + \tfrac{1}{2}\kappa)\hat{a} - \sqrt{\kappa}\hat{a}_{\rm in}. \quad (S31)$$

Following Sec. I A, the measurement signal is

$$\langle\hat{M}\rangle_1 - \langle\hat{M}\rangle_0 = 4|\tilde{g}_{\rm z}|\sin(\varphi - \phi_{\rm h})\tau\cos^2(\tfrac{1}{2}\varphi_{\rm qb})\left\{1 - \frac{2}{\kappa\tau}\left[\cos(\varphi_{\rm qb}) - \cos(\varphi_{\rm qb} + \chi\tau)e^{-\frac{1}{2}\kappa\tau}\right]\right\}, \quad (S32)$$

where as before we note the dispersive-coupling-induced rotation $\varphi_{\rm qb} = 2\arctan(2\chi/\kappa)$. Again as above, the measurement noise is not changed by the dispersive shift. Choosing $\varphi - \phi_{\rm h} = \frac{\pi}{2}$ mod $\pi$, the SNR finally reads

$$\text{SNR}(\chi) = \sqrt{8}\frac{|\tilde{g}_{\rm z}|}{\kappa}\sqrt{\kappa\tau}\cos^2(\tfrac{1}{2}\varphi_{\rm qb})\left\{1 - \frac{2}{\kappa\tau}\left[\cos(\varphi_{\rm qb}) - \cos(\varphi_{\rm qb} + \chi\tau)e^{-\frac{1}{2}\kappa\tau}\right]\right\}. \quad (S33)$$

The residual dispersive coupling reduces the value of the SNR, with the decrease behaving differently at long and short measurement times. At long measurement times, the dispersive coupling reduces the SNR by

$$\text{SNR}(\chi) \simeq \cos^2(\tfrac{1}{2}\varphi_{\rm qb})\text{SNR}(\chi=0) = \frac{\kappa^2}{\kappa^2 + 4\chi^2}\text{SNR}(\chi=0), \qquad \text{for } \tau \gg 1/\kappa. \quad (S34)$$

The SNR is not affected for $\chi \ll \kappa/2$. Interestingly, at short measurement times the SNR is completely independent of the spurious dispersive shift to leading orders

$$\text{SNR}(\chi) \simeq \tfrac{1}{\sqrt{2}}\frac{\tilde{g}_{\rm z}}{\kappa}(\kappa\tau)^{3/2}(1 - \tfrac{1}{6}\kappa\tau), \qquad \text{for } \tau \ll 1/\kappa. \quad (S35)$$

In short, the SNR is not affected by a spurious transverse coupling for short measurement times $\tau \ll 1/\kappa$.

## II. CIRCUIT QED REALIZATION

We now turn to a possible realization of longitudinal coupling in circuit QED. While a lumped circuit is presented in the main text, here we focus on a transmon qubit that is phase-biased by a coplanar waveguide resonator. We recover the lumped element results in the appropriate limit. *We emphasize that the numerical results presented here are obtained for this coplanar realization. For this reason these numerical values differ from, but are compatible with, what is found in the main text.*

### A. Circuit Lagrangian and Hamiltonian

As illustrated in Fig. S1, we consider a transmon qubit coupled to the end of the center conductor of a $\lambda/4$ resonator. To increase the coupling a Josephson junction can be inserted in the center conductor of the resonator at the location of the qubit; see dark region in the resonator's center conductor of Fig. S1. This is reminiscent to the approach to ultra-strongly couple a flux qubit to a resonator discussed in Ref. [S6]. The modelling of the present circuit closely follows that reference and we include the details here for completeness.

The Lagrangian of this circuit, $\mathcal{L} = \mathcal{L}_\mathrm{r} + \mathcal{L}_\mathrm{q} + \mathcal{L}_\mathrm{qr}$, is composed of three parts consisting of the bare resonator $\mathcal{L}_\mathrm{r}$, qubit $\mathcal{L}_\mathrm{q}$ and interaction $\mathcal{L}_\mathrm{qr}$ Lagrangians. From standard circuit theory [S7], the resonator Lagrangian takes the form

$$\left(\frac{2\pi}{\Phi_0}\right)^2 \mathcal{L}_\mathrm{r} = \int_{-L}^{0^-} \left(\frac{C^0}{2}\dot\psi^2(x,t) - \frac{1}{2L^0}(\partial_x\psi(x,t))^2\right)dx + E_{J\mathrm{r}}\cos[\psi(0)] + \frac{C_{J\mathrm{r}}}{2}\dot\psi^2(0). \tag{S36}$$

where $\psi(x)$ is the position-dependant field amplitude inside the resonator. In this expression, we have assumed that the resonator has total length $L$ with capacitance $C^0$ and inductance $L^0$ per unit length. In the single mode limit, we write $\psi(x,t) = \psi(t)u(x)$ where $u(x)$ is the mode envelope [S8]. The Josephson junction in the resonator's center conductor has energy $E_{J\mathrm{r}}$ and capacitance $C_{J\mathrm{r}}$. This junction creates a discontinuity $\psi(0) \neq 0$ in the resonator field that will provide the desired longitudinal interaction. The coupling inductance can be replaced by a SQUID, or SQUID array, without significant change to the treatment [S9]. This Lagrangian was already studied in Refs. [S6, S8].

The transmon qubit is composed of a large shunt capacitor $C_S$ and of Josephson junctions of energies $E_{J1}$ and $E_{J2}$, and capacitances $C_{q1}$ and $C_{q2}$ respectively. In terms of the branch fluxes defined on Fig. S1b), the qubit Lagrangian taking into account the coupling to the resonator is

$$\mathcal{L}_\mathrm{q} + \mathcal{L}_\mathrm{qr} = \left(\frac{\Phi_0}{2\pi}\right)^2 \left[\frac{C_{q1}}{2}\dot\phi_1^2 + \frac{C_{q2}}{2}\dot\phi_2^2 + \frac{C_S}{2}\dot\theta^2\right] + E_{J1}\cos[\phi_1] + E_{J2}\cos[\phi_2]. \tag{S37}$$

With the resonator phase bias $\Delta\psi$ across the coupling junction, the flux quantization around the qubit loop reads $\delta \equiv \phi_1 + \phi_2 = \Phi_\mathrm{x} + \psi(0)$. Defining $\phi_1 = \delta/2 - \theta$ and $\phi_2 = \delta/2 + \theta$, we can write the above as

$$\mathcal{L}_\mathrm{q} + \mathcal{L}_\mathrm{qr} = \left(\frac{\Phi_0}{2\pi}\right)^2 \left[\frac{C_{q1}+C_{q2}}{2}\dot\delta^2 + \frac{C_S+C_{q1}+C_{q2}}{2}\dot\theta^2 + (C_{q2}-C_{q1})\dot\delta\dot\theta\right] + E_{J1}\cos[\delta/2-\theta] + E_{J2}\cos[\delta/2+\theta]. \tag{S38}$$

Assuming the resonator phase-bias to be small $\psi(0) \ll 1$, we find to zeroth-order in $\psi(0)$ the usual Lagrangian of an asymmetric flux biased transmon qubit [S10]

$$\mathcal{L}_q = \left(\frac{\Phi_0}{2\pi}\right)^2 \frac{C_S+C_{q1}+C_{q2}}{2}\dot\theta^2 + E_{J\Sigma}\left[\cos(\Phi_\mathrm{x}/2)\cos\theta - d\sin(\Phi_\mathrm{x}/2)\sin\theta\right], \tag{S39}$$

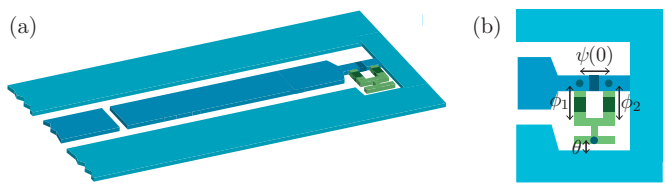

FIG. S1. a) Transmon qubit coupling longitudinally to a resonator. A Josephson junction at the location of the qubit can be used to increase the coupling. b) Closeup of the qubit with the definitions of the branch fluxes used in the text.

where $E_{J\Sigma} = E_{J1} + E_{J2}$ and $d = (E_{J2} - E_{J1})/E_{J\Sigma}$ is the junction asymmetry. Defining $\hat{n}$ to be the conjugate charge to $\hat{\theta}$, the corresponding Hamiltonian is

$$\hat{H}_\text{q} = 4E_C \hat{n}^2 - E_{J\Sigma} \left[ \cos(\Phi_\text{x}/2) \cos\hat{\theta} - d \sin(\Phi_\text{x}/2) \sin\hat{\theta} \right], \tag{S40}$$

with the charging energy $E_C = e^2/2(C_S + C_{q1} + C_{q2})$.

For clarity, we project on the qubit subspace $\{|0\rangle, |1\rangle\}$ where the total Hamiltonian takes the form

$$\hat{H} = \omega_\text{r} \hat{a}^\dagger \hat{a} + K(\hat{a}^\dagger \hat{a})^2 + \frac{\omega_\text{a}}{2} \hat{\sigma}_z + \hat{H}_\text{qr}. \tag{S41}$$

In this expression, $\omega_\text{a}$ is the qubit transition frequency and $K$ the Kerr non-linearity. The latter can be made small and in particular negligible with respect to the photon decay rate $\kappa$. To first-order in $\psi(0)$ and in the same two-level approximation, the interaction Hamiltonian reads

$$\begin{aligned}\hat{H}_\text{qr} &= E_{J\Sigma} \frac{\psi(0)}{2} \left[ \sin(\Phi_\text{x}/2) \cos\hat{\theta} + d \cos(\Phi_\text{x}/2) \sin\hat{\theta} \right] \\ &= g_\text{z}(\hat{a}^\dagger + \hat{a})\hat{\sigma}_z + g_\text{x}(\hat{a}^\dagger + \hat{a})\hat{\sigma}_x,\end{aligned} \tag{S42}$$

where

$$g_\text{z}(\Phi_\text{x}) = -\psi_\text{rms} \frac{E_{J\Sigma}}{4} [m_{11}(\Phi_\text{x}) - m_{00}(\Phi_\text{x})], \tag{S43}$$

$$g_\text{x}(\Phi_\text{x}) = \psi_\text{rms} \frac{E_{J\Sigma}}{2} m_{01}(\Phi_\text{x}). \tag{S44}$$

To obtain these expressions we have used $\psi(0) = u(0)\psi_\text{rms}(\hat{a}^\dagger + \hat{a})$, with $\psi_\text{rms} = \sqrt{4\pi Z_\text{r}/R_K}$. Here $Z_\text{r}$ is the resonator mode impedance and $R_K = h/e^2$ the quantum of resistance. The mode amplitude $u(0)$ at the location of the resonator's junction is related to the participation ratio of the coupling inductance $L_J^{-1} = E_{Jr}(2\pi/\phi_0)^2$ to the total inductance of the resonator mode $L_r$ as $u(0) \approx \eta = L_J/L_r$ [S8]. The lumped element limit discussed in the main text is obtained for $\eta \to 1$. Moreover, we have defined $m_{ij} = \langle i| \sin(\Phi_\text{x}/2) \cos\hat{\theta} + d \cos(\Phi_\text{x}/2) \sin\hat{\theta} |j\rangle$. Eqs. (S44) and (S43) are calculated numerically by diagonalizing the transmon Hamiltonian; see full lines in Fig. 3(b).

### B. Numerical evaluation of the coupling strength

We now present a set of possible parameters for this circuit. *Is is very important to emphasize that these numbers relate to the coplanar architecture discussed here and not to the lumped-element version found in the main text.* We consider a $\lambda/4$ resonator of characteristic impedance $Z^0 \sim 50$ $\Omega$, total length 3.2 mm and capacitance per unit length $C^0 = 0.111$ nF/m and inductance per unit length $L^0 = 0.278$ $\mu$H/m. The coupling inductance consists of an array of 13 josephson Junctions with a total Josephson energy $E_{Jr}/h = 420$ GHz. Following Ref. [S8], we find the first resonator mode to be at frequency $\omega_\text{r}/2\pi = 9.99$ GHz, with a characteristic mode impedance $Z_\text{r} = 44.7$ $\Omega$. The participation ratio of the array is found to be $\eta \sim 0.53$. With these parameters, the resonator bias on the qubit is $\psi_\text{rms} \approx 0.079$.

The transmon Josephson and charging energies are $E_{J\Sigma}/h = 20$ GHz and $E_c/h = 0.3$ GHz respectively, and the asymmetry $d = 0.02$ [S11]. These parameters yield $\omega_\text{a}/2\pi = 6.6$ GHz when evaluated at the flux sweet-spot $\Phi_\text{x} = 0$. The magnetic flux is modulated around this flux value with a maximum excursion of $\widetilde{\Phi}_\text{x}/2\pi = 0.1$. At the sweet-spot, the qubit-resonator is $\Delta = \omega_\text{a}(0) - \omega_\text{r} \approx 2\pi \times 3.4$ GHz. The longitudinal coupling $\widetilde{g}_z$ and the various spurious interactions found numerically using these numbers are summarized in Table I. In particular, we find $\widetilde{g}_z/2\pi = 21.5$ MHz and a small transverse coupling resulting in a negligible dispersive shift $\chi_\text{x} = g_\text{x}^2/\Delta \sim 13$ KHz. The maximum change in qubit frequency is $\omega_\text{a}(0) - \omega_\text{a}(\widetilde{\Phi}_\text{x}) \sim 2\pi \times 170$ MHz.

The array of $N$ junctions allows for strong longitudinal coupling while reducing the resonator nonlinearity. Compared to a single junction of equal energy, the non-linear Kerr effectively decreases as $K = K_1/N^2$ [S12] and from the parameters above we get $K/2\pi = 48$ KHz using 13 junctions.

TABLE I. Longitudinal coupling rate $\widetilde{g}_z$ and spurious leading-order and second-order couplings for a transmon coupled to a $\lambda/4$ resonator with a coupling inductance consisting of an array of 13 Josephson junctions. See text for the parameters. The numbers refer to the coplanar architecture, not the lumped element version discussion in the main text.

| Couplings | | Rates ($2\pi\times$ MHz) |
|---|---|---|
| Longitudinal [Eq. (S46)] | $\widetilde{g}_z$ | 21.5 |
| Transverse [Eq. (S47)] | $g_x$ | 6.56 |
| Dispersive | $g_x^2/\Delta$ | 0.013 |
| Non-linear dispersive [Eq. (S54)] | $2\bar{\Lambda}_z$ | 2.6 |
| Non-linear transverse [Eq. (S55)] | $\bar{\Lambda}_x$ | 0 |
| Kerr nonlinearity | $K$ | 0.048 |
| Qubit frequency shift (from flux-drive) [Eq. (S52)] | $\epsilon_q^2/\Delta$ | $4.4\times 10^{-5}$ |

### C. Asymptotic expression for the coupling strengths

To obtain asymptotic expressions for the qubit-resonator coupling, we consider the transmon as a weakly anharmonic oscillator [S10]. In this situation, $\hat{\theta} \approx \left[\frac{2E_C}{E_J(\Phi_x)}\right]^{1/4}(\hat{b}^\dagger + \hat{b})$ and

$$m_{i,j\neq i} \approx d\left(\frac{2E_C}{E_J(\Phi_x)}\right)^{1/4} \cos(\Phi_x/2) \langle i|(\hat{b}^\dagger + \hat{b})|j\rangle, \quad m_{ii} \approx -\left(\frac{2E_C}{E_J(\Phi_x)}\right)^{1/2} \frac{\sin(\Phi_x/2)}{2} \langle i|(\hat{b}^\dagger + \hat{b})^2|i\rangle. \quad \text{(S45)}$$

Restricting to the $\{|0\rangle, |1\rangle\}$ subspace and using Eqs. (S44) and (S43), we find the asymptotic expressions

$$g_z \approx -\frac{E_{J\Sigma}}{2}\left[\frac{2E_C}{E_J(\Phi_x)}\right]^{1/2}\sqrt{\frac{\pi Z_r}{R_K}}\sin(\Phi_x/2)\,\eta, \quad \text{(S46)}$$

$$g_x \approx dE_{J\Sigma}\left[\frac{2E_C}{E_J(\Phi_x)}\right]^{1/4}\sqrt{\frac{\pi Z_r}{R_K}}\cos(\Phi_x/2)\,\eta. \quad \text{(S47)}$$

These correspond to Eqs. (5) and (6) of the main text in the lumped-element limit $\eta \to 1$.

### D. Upper bounds on the coupling strength

An upper bounds for the longitudinal coupling is obtained by expressing the $g_z$ in units of the qubit frequency. Using Eq. (S46), we find

$$\frac{g_z}{\omega_a} \approx \frac{\Phi_x}{8}\sqrt{\frac{\pi Z_r}{R_K}}\eta. \quad \text{(S48)}$$

For $Z_r \sim 50\,\Omega$ resonator, maximal coupling is reached in the lumped-element limit of the oscillator where $\eta \to 1$ and we get $g_z/\omega_a \sim 0.006$, or about 4 times the coupling we obtained in our previous example giving $g_z/2\pi \sim 42.4$ MHz.

Even larger values of $g_z$ can be achieved by using lumped LC-circuit comprised of a superinductance of large impedance $Z_r \sim R_K/4$ [S13]. For participation ratios in the range $\eta \sim [10^{-2}, 1]$, the coupling is enhanced by a factor of $\sim 10$ to $g_z/\omega_a = \Phi_x \eta \sqrt{\pi}/8 \sim [7\times 10^{-4}, 7\times 10^{-2}]$. While the previous circuit model would have to be refined to take the large phase bias $\Delta\psi$ into account, it is safe to say that the larger and more compact the coupling inductance is, the stronger the longitudinal coupling will be.

## III. SPURIOUS COUPLINGS AND IMPERFECTIONS

### A. Effect of the flux drive

From the first term in Eq. (S38), an oscillating external magnetic flux $\Phi_x(t) = \widetilde{\Phi}_x \cos(\omega_r t + \varphi)$ at frequency $\omega_d$ leads to an effective voltage drive on the resonator

$$\mathcal{L}_{\rm r,drive} = \left(\frac{\Phi_0}{2\pi}\right)^2 (C_{q1} + C_{q2})\dot{\Phi}_x \dot{\psi}(0). \tag{S49}$$

This term leads to an effective drive on the resonator $\hat{H}_{r,d} = \epsilon_r e^{-i(\omega_r t + \varphi)} \hat{a}^\dagger + {\rm h.c.}$ of drive amplitude $\epsilon_r$

$$\epsilon_r = \frac{\hbar\omega_r}{8E_{CJ}} \psi_{\rm rms}\, \Phi_x\, \omega_d, \tag{S50}$$

where $E_{CJ} = e^2/[2(C_{q1} + C_{q2})]$. With the above circuit parameters, we find $\epsilon_r/2\pi \sim 10$ MHz. If desired, the effect of this drive can be cancelled by an additional on the input port of the resonator. Otherwise, this simply leads to an additional qubit-state *independent* displacement of the cavity field that does not affect the signal-to-noise ratio.

In the same way, flux modulation also directly drives the qubit. This is caused by the last charging energy term of Eq. (S38) and yields

$$\hat{H}_{q,d} = \left(\frac{\Phi_0}{2\pi}\right)^2 (C_{q2} - C_{q1})\dot{\Phi}_x \hat{q}, \tag{S51}$$

From the asymptotic expression $\hat{q} \approx i(E_J/32E_C)^{1/4}(b^\dagger - b)$ we get $\hat{H}_{q,d} = \epsilon_q e^{-i(\omega_r t + \varphi)} \hat{b}^\dagger + {\rm h.c.}$ where

$$\epsilon_q \approx \frac{d}{4} \frac{C_J}{C_\Sigma} \left(\frac{E_J}{32E_C}\right)^{1/4} \Phi_x \omega_r \tag{S52}$$

is identically zero for symmetric junctions ($d = 0$). With the above parameters and $C_J/C_\Sigma = 0.01$, we find that $\epsilon_q/2\pi \approx 0.36$ MHz. Given the strong qubit-resonator detuning of several GHz, this is however of no consequences.

### B. Higher-order interaction terms

To second order in $\Delta\psi$ in Eq. (S38) yields the additional interactions

$$\hat{H}_{\rm qr}^{(2)} = (\hat{a}^\dagger + \hat{a})^2 (\Lambda_x \hat{\sigma}_x + \Lambda_z \hat{\sigma}_z), \tag{S53}$$

where

$$\Lambda_z = \psi_{\rm rms}^2 \frac{E_{J\Sigma}}{16} \left(\langle 0| \hat{cc} - d\, \hat{ss} |0\rangle - \langle 1| \hat{cc} - d\, \hat{ss} |1\rangle\right), \tag{S54}$$

$$\Lambda_x = \psi_{\rm rms}^2 \frac{E_{J\Sigma}}{8} \langle 1| [\hat{cc} - d\, \hat{ss}] |0\rangle, \tag{S55}$$

and where we have defined $\hat{cc} = \cos(\Phi_x/2)\cos\hat{\theta}$ and $\hat{ss} = \sin(\Phi_x/2)\sin\hat{\theta}$.

Under flux modulation, $\Lambda_k(t) = \bar{\Lambda}_k + \widetilde{\Lambda}_k \cos(\omega_r t + \varphi_r)$ with $k = x, z$. With the RWA this leads to

$$\hat{H}_{\rm qr}^{(2)} \approx (2\hat{a}^\dagger \hat{a} + 1)(\bar{\Lambda}_x \hat{\sigma}_x + \bar{\Lambda}_z \hat{\sigma}_z). \tag{S56}$$

For $d$ small and flux modulations around $\Phi_x = 0$ then $\bar{\Lambda}_x = 0$. On the other hand, the dispersive-like interaction of amplitude $\chi_z = 2\bar{\Lambda}_z$ has a magnitude of $\chi_z/2\pi \approx 2.6$ MHz for the above circuit parameters. In the lumped-element limit with $Z_r = 50$ $\Omega$, we rather find $\chi_z/2\pi = 5.3$ MHz. Since this dispersive-like interaction originates from a second-order correction, we always find $g_z/\chi_z \approx 10$. In practice, we can therefore chose $\chi_z < \kappa$ such that this spurious coupling will not affect the SNR at short integration times.

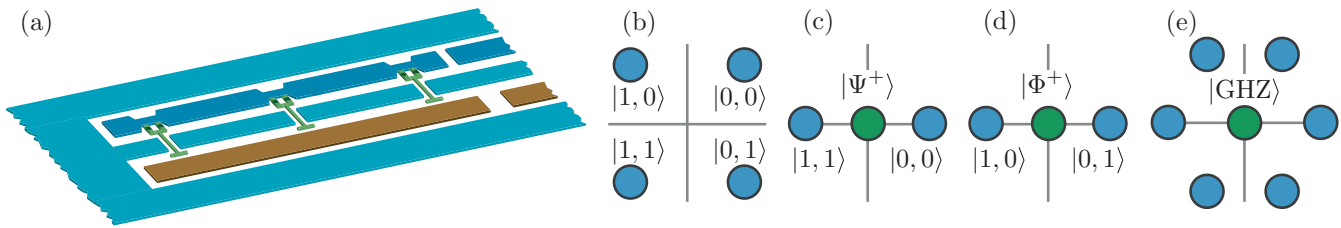

FIG. S2. (a) Multi-qubit architecture consisting of qubits longitudinally coupled to a readout resonator (dark blue) and transversally coupled to a bus resonator (brown). (b)–(e) Phase space representation of the pointer states for two and three qubits. (b) Two-qubit joint readout ($\varphi_j = j\pi/2$), (c-d) Two-qubit entanglement by measurement. The green circles correspond to the Bell states $|\Psi^+\rangle = \frac{1}{\sqrt{2}}[|0,1\rangle + |1,0\rangle]$ and $|\Phi^+\rangle = \frac{1}{\sqrt{2}}[|0,0\rangle + |1,1\rangle]$, obtained for $\varphi_j = 0$ and $\varphi_j = j\pi$, respectively. (e) The green circle corresponds to the GHZ state $|\text{GHZ}\rangle = \frac{1}{\sqrt{2}}[|0,0,0\rangle + |1,1,1\rangle]$. This configuration is obtained for $\varphi_j = j2\pi/3$.

## IV. MULTIPLE QUBIT READ-OUT IN (X+Z) CQED

Fig. S2(a) illustrates a possible multi-qubit architecture: qubits (green) are longitudinally coupled to a readout resonator (dark blue) and transversally coupled to a bus resonator (brown). The readout resonator is used only when the longitudinal couplings are modulated and individual flux control for each qubit is possible with separated flux lines. The system Hamiltonian reads

$$\hat{H} = \omega_{\text{rz}} \hat{a}_z^\dagger \hat{a}_z + \omega_{\text{rx}} \hat{a}_x^\dagger \hat{a}_x + \sum_j \tfrac{1}{2}\omega_{aj}\hat{\sigma}_{zj} + \sum_j g_{zj}\hat{\sigma}_{zj}(\hat{a}_z^\dagger + \hat{a}_z) + \sum_j g_{xj}\hat{\sigma}_{xj}(\hat{a}_x^\dagger + \hat{a}_x). \tag{S57}$$

In the interaction picture and neglecting fast-oscillating terms, the longitudinal coupling becomes

$$\tilde{H}_z = \left(\tfrac{1}{2}\tilde{g}_z \sum_j \hat{\sigma}_{zj} e^{-i\varphi_j}\right)\hat{a}_z + \text{h.c.} \tag{S58}$$

The position of the qubits and the coupling inductances can be adjusted to get equal longitudinal coupling strengths $|g_{zj}|$. By properly choosing the phases $\varphi_j$ it is possible to operate in the joint qubit readout mode or in the entanglement by measurement mode. Examples for 2 and 3 qubits are shown in Fig. S2(b)–(e): joint measurement of two qubits, synthesis of Bell states and 3-qubit GHZ state by measurement, cf. caption.



\end{document}